\documentclass[prl,twocolumn]{revtex4}
\usepackage{graphicx}
\usepackage{amsfonts,amsmath,amssymb,float}
\usepackage{bm,dsfont}
\usepackage{color}
\usepackage{bbm}
\usepackage{longtable}
\usepackage[dvips]{epsfig}
\usepackage{amsmath,amssymb,lscape,float}
\usepackage{hyperref}
\usepackage{listings}

\begin{document}
\title{Quantum circuits for digital quantum simulation of nonlocal electron-phonon coupling
}

\author{Vladimir M. Stojanovi\'c}
\email{vladimir.stojanovic@physik.tu-darmstadt.de}
\affiliation{Institut f\"{u}r Angewandte Physik, Technical
University of Darmstadt, D-64289 Darmstadt, Germany}

\date{\today}

\begin{abstract}
Motivated by the compelling need to understand the nonequilibrium dynamics of small-polaron 
formation following an electron-phonon interaction quench, in this work we propose a digital 
quantum simulator of a one-dimensional lattice model describing an itinerant fermionic excitation 
(e.g. an electron) nonlocally coupled to zero-dimensional bosons (e.g. Einstein-type phonons). 
Quantum circuits implementing the dynamics of this model, which includes Peierls- and breathing-mode-type 
excitation-boson interactions, are designed here, their complexity scaling linearly with the system 
size. A circuit that generates the natural initial (pre-quench) state of this system -- a bare-excitation 
Bloch state, equivalent to a $W$ state of a qubit register -- is also presented. To facilitate 
comparisons with the proposed simulator, once experimentally realized, the system dynamics are 
also evaluated numerically and characterized through the Loschmidt echo and various correlation 
functions.
\end{abstract}

\maketitle

Inspired by Feynman's pioneering idea that a quantum system could efficiently be 
simulated using a device that is also governed by quantum-mechanical laws~\cite{feynmansim}, 
the field of quantum simulation is at the current frontier of physics research~\cite{Altman++:21}.
While analog quantum simulators of many-body systems were intensively studied 
already a decade ago~\cite{Georgescu+:14}, the ongoing spate of activity in the realm 
of {\em digital quantum simulation} (DQS)~\cite{Lloyd:96,Zalka:98,Georgescu+:14} is 
mainly motivated by the considerable progress in the development of quantum
hardware~\cite{PreskillNISQ:18} achieved in recent years~\cite{Wendin:17,Bruzewicz+:19,MorgadoWhitlock:21}.
In particular, owing to the relatively modest resources required for their execution, 
quantum algorithms for simulating fermionic systems have as yet been most extensively 
studied~\cite{Abrams+Lloyd:9799,Somma+:02,Bravyi+Kitaev:02,Whitfield+:11,Raeisi+:12,Wecker+:15,Barends+:15,
Babbush+:18,Reiner+:18,Jiang+:18,Rubin+:21,McArdle+:20}. On the other hand, digital simulators 
of bosonic (or fermion-boson) systems have heretofore received much less attention~\cite{Mezzacapo+:12,
Macridin+:18,Macridin+:22,DiPaolo+:20,Fitzpatrick+:21,Miessen+:21}. 

Electron-phonon (e-ph) coupling is pervasive in solid-state systems~\cite{GiustinoRMP:17}, being 
arguably the most important interaction mechanism that involves both bosons and fermions. In particular, 
one common physical situation in narrow-band semiconductors and insulators entails an itinerant 
excitation (electron, hole) experiencing a short-ranged, nonpolar interaction with the 
host-crystal lattice vibrations. Such phenomena are typically described within the 
framework of the Holstein model~\cite{Holstein:59,Wellein+Fehske:97}, the strong-coupling 
regime of which captures the formation of a heavily-dressed quasiparticle known as 
{\em small polaron} (SP)~\cite{Emin:82}. While this model only takes into account local coupling 
of the excitation density and an Einstein-phonon displacement, the need to describe 
complex electronic materials~\cite{GiustinoRMP:17} has prompted investigations of 
nonlocal e-ph interactions~\cite{Hannewald++:04,Stojanovic+:04,Shneyder+:20,Stojanovic++:10}. 
This line of research has even led to the discovery of sharp SP ground-state transitions of 
level-crossing type~\cite{Stojanovic+Vanevic:08,Sous+:17,Stojanovic+:14} in certain models
with strongly momentum-dependent e-ph interactions~\cite{Stojanovic:20}. 

While several analog simulators of coupled e-ph models were proposed in the past~\cite{Stojanovic+:12,Herrera+:13,
Mei+:13,Stojanovic+:14,Stojanovic+Salom:19,Nauth+Stojanovic:23}, only the Holstein model was addressed in the DQS 
context~\cite{Mezzacapo+:12,Macridin+:18,Jaderberg+:22}. Aiming to change this state of affairs, this paper presents 
DQS of a one-dimensional (1D) lattice model describing an itinerant fermionic excitation (e.g. an electron) 
coupled to dispersionless bosons (e.g. Einstein-type phonons) through breathing-mode-~\cite{Slezak++:06} 
and Peierls-type~\cite{Stojanovic+Vanevic:08} interactions. Quantum circuits emulating these 
short-ranged yet nonlocal interactions are presented here, with their complexity scaling linearly 
with the system size.

In contrast to the existing works on DQS of coupled e-ph models~\cite{Mezzacapo+:12,Macridin+:18}, 
which solely discuss the already well-understood ground-state properties of Holstein-type SPs~\cite{Wellein+Fehske:97}, 
the present work is chiefly concerned with the nonequilibrium SP physics; the ground-state properties will 
also be discussed, but mainly for benchmarking purposes. The current understanding of 
the latter, especially of the complex process of SP formation following an e-ph interaction quench (i.e., a 
sudden switching-on of the e-ph coupling) is by no means satisfactory~\cite{Ku+Trugman:07,Fehske+:11}. In 
principle, recent advances in ultrafast time-resolved spectroscopies~\cite{UltrashortExp} opened the possibilities 
to investigate such processes experimentally. Yet, many important questions remain unanswered due to the 
impossibility of fully separating the effects of e-ph coupling from the other interaction mechanisms in 
solid-state systems. Thus, the incentive to explore the quench dynamics of SP formation within a 
well-controlled DQS setting -- as attempted in this paper -- is compelling. 

\textit{Model Hamiltonian}.-- The model under consideration envisions a spinless-fermion excitation 
with the hopping amplitude $t_e$ on a 1D lattice where each site hosts dispersionless bosons (or, 
equivalently, a zero-dimensional harmonic oscillator). The noninteracting part of the system Hamiltonian 
$\hat{H}_{\textrm{sys}}$ consists of the excitation kinetic-energy- and free-boson terms 
(hereafter $\hbar=1$):
\begin{equation}\label{H_0}
\hat{H}_{0} = -t_e\sum_n (\hat{c}^\dagger_{n+1}\hat{c}_n + \mathrm{H.c.}) 
+ \omega_{\textrm b}\sum_n \hat{b}^\dagger_n \hat{b}_n \:.
\end{equation}
Here $\hat{c}^\dagger_n$ ($\hat{c}_n$) creates (destroys) a spinless-fermion excitation at site 
$n$ ($n=1,\ldots,N$) and $\hat{b}^\dagger_n$ ($\hat{b}_n$) a boson with frequency $\omega_{\textrm b}$ 
at the same site. The interacting (e-b) part of $\hat{H}_{\textrm{sys}}$ is given 
by $\hat{H}_{\textrm{e-b}}=\hat{H}_{\textrm{P}}+\hat{H}_{\textrm{B}}$, where
\begin{eqnarray}\label{H_eb}
\hat{H}_{\textrm{P}} &=& \alpha_{\textrm{P}}
\sum_n(\hat{c}^\dagger_{n+1}\hat{c}_n+\mathrm{H.c.})(\hat{x}_{n+1}-\hat{x}_n) \:, \nonumber\\
\hat{H}_{\textrm{B}} &=& \alpha_{\textrm{B}}
\sum_n \hat{c}^\dagger_n \hat{c}_n(\hat{x}_{n-1}-\hat{x}_{n+1})\:,
\end{eqnarray}
and $\hat{x}_n\equiv \xi_0(\hat{b}_n^\dagger + \hat{b}_n)$ is the displacement 
operator of the oscillator at site $n$, with $\xi_0$ being its zero-point length.
$\hat{H}_{\textrm{P}}$ accounts for the lowest-order dependence of the effective 
excitation-hopping amplitude between two adjacent lattice sites on the respective displacements 
(Peierls-type coupling)~\cite{Stojanovic+:04,Stojanovic+Vanevic:08}, while $\hat{H}_{\textrm{B}}$ 
describes the antisymmetric coupling of the excitation density on a given site with 
the displacements at the two adjacent sites (breathing-mode-type coupling)~\cite{Slezak++:06}. 
As a manifestation of the discrete translational symmetry, the Hamiltonian 
$\hat{H}_{\textrm{sys}}=\hat{H}_{0}+\hat{H}_{\textrm{P}}+\hat{H}_{\textrm{B}}$ 
commutes with the total quasimomentum operator $\hat{K}=\sum_{k}k\:\hat{c}_{k}^{\dagger}
\hat{c}_{k}+\sum_{q}q\:\hat{b}_{q}^{\dagger}\hat{b}_{q}$, where $\hat{c}_{k}^{\dagger}$ and 
$\hat{b}_{q}^{\dagger}$ are the respective momentum-space counterparts of the operators 
$\hat{c}^\dagger_n$ and $\hat{b}_n^\dagger$.

\textit{Quench dynamics from DQS}.-- The system dynamics following an e-b interaction quench at $t=0$ 
are considered here within the DQS framework. Use will be made of the first-order Trotter-Suzuki-type 
decomposition~\cite{Trotter:59,Suzuki:76}. For a Hamiltonian $\hat{H}=\sum_l\hat{H}_l$ the latter
approximates $\exp(-i\hat{H}t)$ by $(\prod_l e^{-i\hat{H}_l t/n_s})^{n_s}$, with the corresponding error 
being upper bounded by $\mathcal{O}(Nt^2/n_s)$, where $N$ is the number of qubits and $n_s$ that of time 
steps~\cite{Lloyd:96}. To this end, the time-evolution operator $\hat{U}_{\textrm{sys}}
\equiv\exp(-i\hat{H}_{\textrm{sys}}t)$ of the system at hand is represented through quantum circuits 
that emulate the dynamics inherent to the Hamiltonians $\hat{H}_{0}$, $\hat{H}_{\textrm{P}}$, and 
$\hat{H}_{\textrm{B}}$.  

In the DQS context one switches from spinless-fermion- to qubit degrees of freedom via the 
Jordan-Wigner transformation (JWT)~\cite{Jordan+Wigner:28}, which
assumes an ordering of the qubits and maps the fermionic operators $\hat{c}^\dagger_n$ and 
$\hat{c}_n$ to the (qubit) Pauli operators $X_n,Y_n,Z_n$ according to $2\hat{c}^\dagger_n
\rightarrow (X_n-iY_n)Z_1\ldots Z_{n-1}$ and $2\hat{c}_n\rightarrow (X_n+iY_n)Z_1\ldots Z_{n-1}$. 
The state $|1_n\rangle$ of qubit $n$ corresponds to a spinless-fermion occupying the 
lattice site $n$, while $|0_n\rangle$ represents an unoccupied site $n$.

The natural initial (pre-quench) state $|\psi_{t=0}\rangle$ of the system at hand is the ground state 
of the Hamiltonian $\hat{H}_{0}$, i.e. the bare-excitation Bloch state at zero quasimomentum $|\Psi_{k=0}
\rangle\equiv c^{\dagger}_{k=0}|0\rangle_{\textrm{e}}\otimes|0\rangle_{\textrm{b}}$. Using the JWT
it is straightforward to establish that $|\Psi_{k=0}\rangle=|W_N\rangle\otimes|0\rangle_{\textrm{b}}$, 
where $|W_N\rangle$ is an $N$-qubit $W$ state~\cite{StojanovicPRL:20,StojanovicPRA:21}:
\begin{equation}\label{defWstate}
|W_N\rangle = N^{-1/2}\sum_{n=1}^{N}|0\ldots 1_n\ldots 0\rangle \:.
\end{equation} 

\begin{figure}[t]
\includegraphics[width=0.9\linewidth]{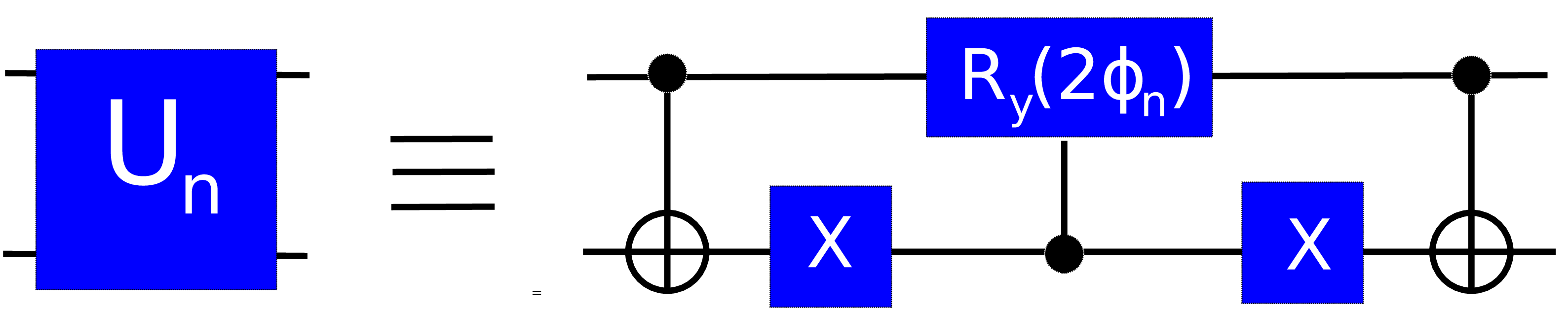}
\caption{\label{fig:CircuitTsfU}(Color online) Quantum-circuit representation of 
the unitaries $U_{n}(\phi_n)$, where $R_y(2\phi_n)\equiv\exp(-i\phi_n Y)$ 
stands for a $y$-axis rotation by an angle of $2\phi_n$ ($n=1,\ldots,N$).}
\end{figure}

\textit{$W$-state generating circuit}.-- In what follows, a quantum circuit is proposed that generates 
the state in Eq.~\eqref{defWstate} when acting on an $N$-qubit register. The circuit constitutes a digital 
counterpart of an analog scheme in which $W$ states are generated through sequential entanglement 
of an auxiliary qubit $0$ with each of the qubits $1,2,\ldots, N$~\cite{Schoen+:07}; it is assumed that 
the whole $(N+1)$-qubit system is initalized in the state $|0_0\rangle\otimes|0\rangle^{\otimes N}$. To 
create entanglement between qubits $0$ and $n$ ($n=1,\ldots,N$), one considers the following parameterized 
unitary operation:
\begin{eqnarray}\label{defUnitary}
 U_{n}(\phi_n)&=&|0_0 1_n\rangle\langle 0_0 1_n|+|1_0 0_n\rangle\langle 1_0 0_n| \nonumber\\
 &+& \cos\phi_n(|0_0 0_n\rangle\langle 0_0 0_n|+|1_0 1_n\rangle\langle 1_0 1_n|)  \\
 &+& \sin\phi_n(|1_0 1_n\rangle\langle 0_0 0_n|-|0_0 0_n\rangle\langle 1_0 1_n|) \:. \nonumber
\end{eqnarray}
For each choice of $\phi_n$, $U_n$ generates a superposition of $|0_0 0_n\rangle$ 
and $|1_0 1_n\rangle$, acting trivially on the remaining $N-1$ qubits. The circuit
representation of $U_n$, which entails two single-qubit (NOT) gates and three two-qubit 
gates [one controlled-$y$ rotation- and two controlled-NOT (CNOT) gates], is shown in 
Fig.~\ref{fig:CircuitTsfU}.

\begin{figure}[b!]
\includegraphics[width=0.85\linewidth]{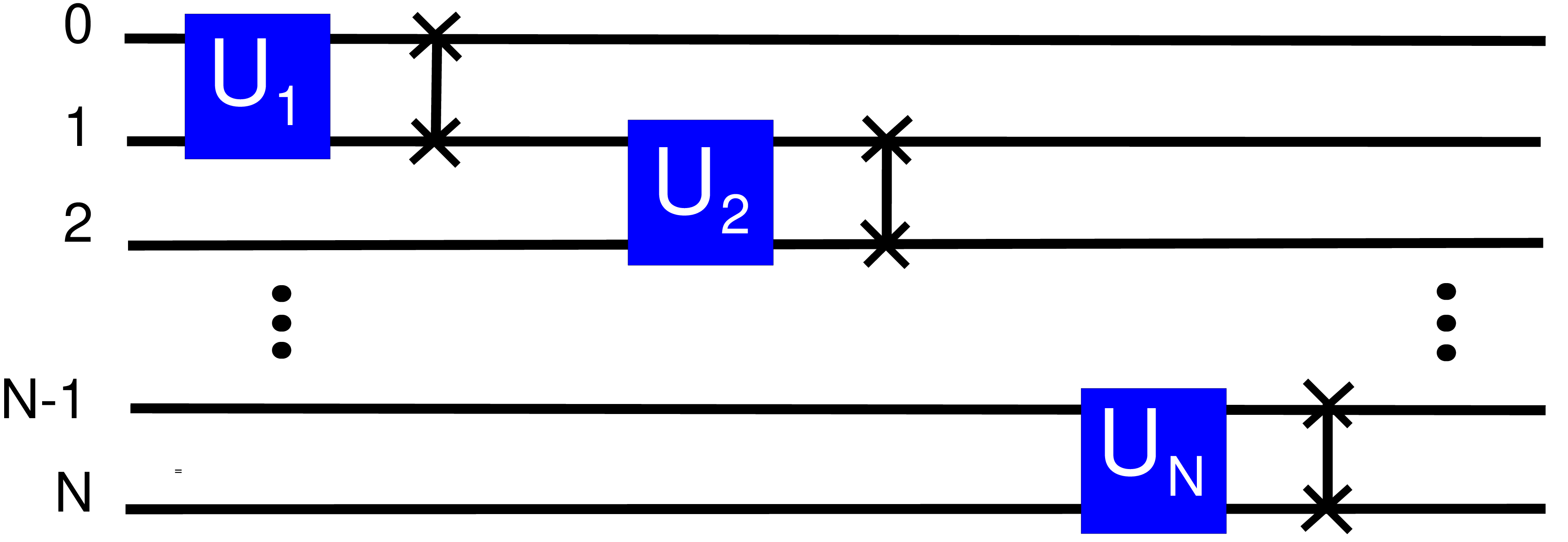}
\caption{\label{fig:WstateCircuit}(Color online) Quantum circuit generating the 
inital $W$-type state of the system when acting on a $(N+1)$-qubit register 
initialized in the state $|0\rangle\otimes|0\rangle^{\otimes N}$.}
\end{figure}

The chosen $W$-state preparation scheme envisions applying $U_{n}(\phi_n)$ for $n=1,\ldots,N$ 
sequentially on the initial state. The first step in this scheme gives rise to 
$U_{1}(\phi_1)|0_0 0_1\rangle = \cos\phi_1|0_0 0_1\rangle + \sin\phi_1|1_0 1_1\rangle$. 
While subsequent application of the unitaries $U_{n}(\phi_n)$ for $n\ge 2$ engenders 
a broad family of entangled states, it is straightforward to show that for one special choice
of $\phi_n$ one obtains the final state $|1\rangle\otimes|W_N\rangle$, where $|W_N\rangle$ 
is given by Eq.~\eqref{defWstate}. This choice is described by the 
conditions $\prod_{n=1}^{N}\cos\phi_n=0$ and $\cos\phi_n|\sin\phi_{n+1}|=|\sin\phi_n|$,
which yield the solution $\sin\phi_n=(N+1-n)^{-1/2}$. The circuit generating the desired 
$W$ state in a 1D qubit array using a sequence of alternating $U_{n}$ and SWAP operations 
is illustrated in Fig.~\ref{fig:WstateCircuit}. 

Based on the circuit representation of $U_n$ 
(cf. Fig.~\ref{fig:CircuitTsfU}) and the fact that a SWAP operation can be realized using three 
CNOT gates~\cite{NielsenChuangBook}, one concludes that $\mathcal{O}(N)$ gates are required 
for generating the state $|W_N\rangle$ using this scheme.

\textit{Boson-state encoding}.-- One of the prerequisites for the envisioned 
DQS is the capability to encode bosonic states on qubits. To this end, use will be made of a recently 
proposed method based on the idea of controlled truncation of bosonic Hilbert spaces~\cite{Macridin+:18}, 
whose rigorous foundation rests on a generalized Nyquist-Shannon sampling theorem~\cite{Macridin+:22}. 
In the case of zero-dimensional oscillators, the number $n_q$ of required qubits per lattice site in this 
approach scales as $\log[\log(1/\epsilon)]$, where $\epsilon$ is the desired accuracy~\cite{Macridin+:18}. 
In keeping with Ref.~\cite{Macridin+:18}, in what follows the states $|x_n\rangle$ of the oscillator 
on site $n$ are encoded in such a way that the index $r$ of each qubit within the register of $n_q$ qubits 
is defined by the binary representation $x_n=\sum_{r=0}^{n_q-1}x^{r}_{n}2^{r}$ of $x_n$ ($x^{r}_{n}=0,1$). 
Finally, the dimensionless ($\xi_0\rightarrow 1$) coordinate and momentum operators in the $2^{n_q}$-dimensional 
truncated Hilbert space of the oscillator $n$ will be denoted by $\tilde{x}_n$ and $\tilde{p}_n$.

In particular, quantum circuits pertaining to the cooordinate part of the free boson Hamiltonian, i.e. 
the time-evolution operator $e^{-i\theta \hat{x}_n^2}$, were proposed in Ref.~\cite{Macridin+:18}. They
involve controlled phase-shift gates $T(\theta_r)\equiv \textrm{diag}(1,e^{-i\theta_r})$ acting on each 
qubit in the boson register, with $\theta_r\equiv 2^r \theta$. On the other hand, the realization of 
circuits corresponding to $e^{-i\theta \hat{p}_n^2}$ requires one to first implement a quantum Fourier 
transform from $\{|p_n\rangle, n=1,\ldots,N\}$ to $\{|x_n\rangle, n=1,\ldots,N\}$, then the same type 
of circuits as for $e^{-i\theta \hat{x}_n^2}$, and, finally, carry out the inverse quantum Fourier 
transform~\cite{Zalka:98}. 

\textit{Circuits for $e^{-i\theta\hat{H}_{\textrm{B}}}$ and $e^{-i\theta\hat{H}_{\textrm{P}}}$}.-- Quantum 
circuits~\cite{NielsenChuangBook} representing the interaction terms $\hat{H}_{\textrm{B}}$ and $\hat{H}_{\textrm{P}}$ 
are presented in the following. They partiall inherit the structure of purely fermionic circuits~\cite{Somma+:02,
Whitfield+:11,Wecker+:15} which are expressed in terms of the Hadamard gates $H_n$, the $z$-rotation gates 
$R_{z,n}(\theta)\equiv\exp(-i\theta Z_n/2)=\textrm{diag}(e^{i\theta/2},e^{-i\theta/2})$, and the phase-shift
gates $T_n(\theta)$ ($n=1,\ldots,N$). The boson (oscillator) states define the attendant rotation- and phase-shift 
angles. 

\begin{widetext}

\begin{figure}[b]
\includegraphics[width=0.975\linewidth]{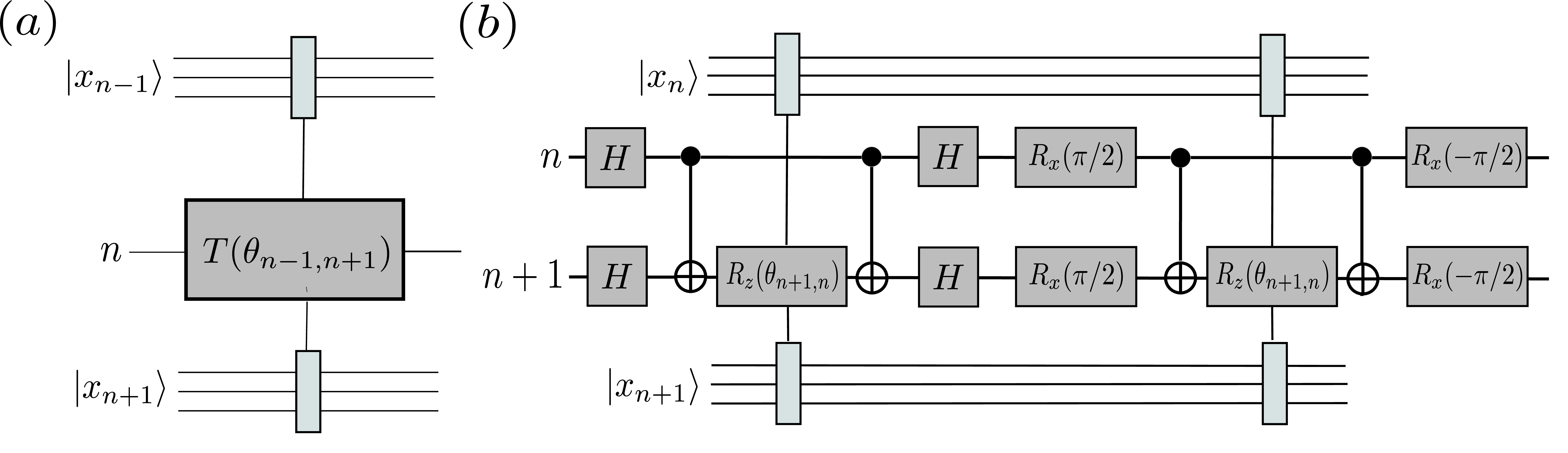}
\caption{\label{fig:CircuitsEPH}(Color online) Quantum circuits corresponding to 
(a) breathing-mode type coupling ($\hat{H}_{\textrm{B}}$), and (b) Peierls-type 
coupling ($\hat{H}_{\textrm{P}}$). The relevant $z$-rotation angles are defined as 
$\theta_{n-1,n+1}\equiv(x_{n-1}-x_{n+1})\theta$ and $\theta_{n+1,n}\equiv(x_{n+1}-x_n)\theta$.}
\end{figure}

\end{widetext}

To find the circuit representation of the time-evolution operator corresponding to the 
Hamiltonian $\hat{H}_{\textrm{B}}$ one makes use of the fact that the JWT maps $\hat{c}^
\dagger_n\hat{c}_n$ into $(\mathbbm{1}-Z_n)/2$...
...where $\theta_{n-1,n+1}\equiv (x_{n-1}-x_{n+1})\theta$.
The resulting circuit implementing $e^{-i\theta\hat{H}_{\textrm{B}}}$ is shown in 
Fig.~\ref{fig:CircuitsEPH}(a).

Recalling that JWT transforms $\hat{c}^\dagger_{n+1}\hat{c}_n+\hat{c}^\dagger_{n}\hat{c}_{n+1}$ 
into $(X_n X_{n+1}+Y_{n}Y_{n+1})/2$ and using a Trotter-type decomposition, the time-evolution 
operator corresponding to the Hamiltonian $\hat{H}_{\textrm{P}}$ can be approximated as $e^{-i\theta
\hat{H}_{\textrm{P}}}\approx U_{Y}(\theta)U_{X}(\theta)$, where
\begin{eqnarray}\label{UxUy}
U_{X}(\theta)&\equiv& e^{-i\frac{\theta}{2}X_n X_{n+1}(\tilde{x}_{n+1}-\tilde{x}_n)} 
\nonumber \:,\\
U_{Y}(\theta)&\equiv& e^{-i\frac{\theta}{2}Y_n Y_{n+1}(\tilde{x}_{n+1}-\tilde{x}_n)}\:.
\end{eqnarray}
By making use of the basis-change transformations $H_n Z_n H_n=X_n$ and $R_{x,n}(-\pi/2)Z_n R_{x,n}
(\pi/2)=Y_n$ and the fact that for an arbitrary unitary $U$ and an analytic operator function $f(A)$ 
it holds that $Uf(A)U^{-1}=f(UAU^{-1})$, one straightforwardly demonstrates that 
\begin{eqnarray}\label{UxUyExpr}
U_{X}(\theta) &=& H_{n}H_{n+1}e^{-i\frac{\theta}{2}Z_n Z_{n+1}(\tilde{x}_{n+1}
-\tilde{x}_n)}H_{n}H_{n+1} \nonumber \:,\\
U_{Y}(\theta) &=& R_{x,n}(-\pi/2)R_{x,n+1}(-\pi/2)e^{-i\frac{\theta}
{2}Z_n Z_{n+1}(\tilde{x}_{n+1}-\tilde{x}_n)}\nonumber \\
&\times& R_{x,n}(\pi/2)R_{x,n+1}(\pi/2)  \:.
\end{eqnarray}
Using the standard circuit representation of the time-evolution operators of the 
type $e^{-i\frac{\theta}{2}Z_n Z_{n+1}}$ with one $R_z(\theta)$ and two CNOT 
gates~\cite{Whitfield+:11}, the operator $e^{-i\frac{\theta}{2}Z_n Z_{n+1}
(\tilde{x}_{n+1}-\tilde{x}_n)}$ can be represented in an analogous fashion, with 
the $z$-rotation angle $\theta_{n+1,n}\equiv(x_{n+1}-x_n)\theta$ determined by the 
states of the oscillators $n$ and $n+1$. 

Based on Eqs.~\eqref{UxUy} and \eqref{UxUyExpr}, 
the resulting circuit implementing $e^{-i\theta\hat{H}_{\textrm{P}}}$ has the form 
illustrated in Fig.~\ref{fig:CircuitsEPH}(b).

\textit{Ground-state evalution}.-- To benchmark the proposed digital simulator
we extract the ground-state properties of the system using quantum phase estimation 
(QPE) algorithm~\cite{Cleve+:98}. These properties can be computed classically -- to 
requisite precision -- using Lanczos-type diagonalization of the total system Hamiltonian. 

The momentum-space form of $H_{\textrm{e-b}}$ reads 
\begin{equation}\label{Heb_ms}
\hat{H}_{\textrm{e-b}}=N^{-1/2}\sum_{k,q}\gamma_{\textrm{e-b}}(k,q)\:
\hat{c}_{k+q}^{\dagger}\hat{c}_{k}(\hat{b}_{-q}^{\dagger}+\hat{b}_{q})\:,
\end{equation}
where the e-b vertex function is given by
$\gamma_{\textrm{e-b}}(k,q)= 2i\omega_{\textrm{b}}\{g_{\textrm{B}}\sin q+g_{\textrm{P}}
[\sin k-\sin(k+q)]\}$, where $g_{\textrm{B}}\equiv\alpha_{\textrm{B}}\xi_0$ and 
$g_{\textrm{P}}\equiv\alpha_{\textrm{P}}\xi_0$ are the respective dimensionless 
coupling strenghts. The most interesting ground-state aspect of such couplings is
the possible occurrence of sharp transitions, i.e. a nonanalyticity of the ground
state energy at a critical coupling strength. It is already known that a Hamiltonian
with Peierls- and breathing-mode-type couplings shows such sharp transition. 
Given that the presence of Peierls-type coupling alone suffices for obtaining a sharp
transition -- while breathing-mode coupling does not have this property (a consequence 
of the fact that its vertex function depends only on $q$) -- it makes sense to use
the effective Peierls-coupling strength $\lambda_{\textrm{P}}$ and the ratio 
$\zeta\equiv g_{\textrm{B}}/g_{\textrm{P}}$ as the two dimensionless parameters 
characterizing e-ph coupling in the system at hand. Recalling that for the most general 
e-b coupling described by the vertex function $\gamma_{\textrm{e-b}}(k,q)$ [cf. Eq.~\eqref{Heb_ms}] 
the corresponding effective e-b coupling strength is given by $\langle|\gamma_{\textrm{e-b}}
(k,q)|^{2}\rangle_{\textrm{BZ}}$, where $\langle\ldots\rangle_{\textrm{BZ}}$ stands 
for the Brillouin zone average, in the special case of Peierls coupling one straightforwardly 
obtains $\lambda_{\textrm{P}} = 2\:g^2_{\textrm{P}}\:\omega_{\textrm{b}}/t_{\rm e}$.

\textit{Dynamics following an interaction quench}.-- The nonequilibrium dynamics of a 
quantum system after an interaction quench at $t=0$ are, generally speaking, best 
characterized by the Loschmidt amplitude~\cite{Heyl:18} 
\begin{equation}\label{LoschmidtAmp}
\mathcal{G}(t)=\langle\psi_{t=0}|\:e^{-i\hat{H}_{\textrm{sys}}t}\:|\psi_{t=0}\rangle \:,
\end{equation}
i.e. the overlap of the initial state $|\psi_{t=0}\rangle$ and its time-evolved counterpart 
$e^{-i\hat{H}_{\textrm{sys}}t}\:|\psi_{t=0}\rangle$ at time $t$. This quantity is closely 
related -- more precisely, up to a Fourier transform to the frequency domain -- to an important 
dynamical response function, namely the momentum-frequency resolved spectral function. 
The squared modulus of $\mathcal{G}(t)$ -- i.e. the survival probability $\mathcal{L}(t)\equiv
|\mathcal{G}(t)|^2$ of the initial state at time $t$ -- is a special case of the quantity known 
as the Loschmidt echo~\cite{Peres:84}. Because the initial state $|\psi_{t=0}\rangle$ (in the 
problem at hand a bare-excitation Bloch state) is, generally speaking, not an eigenstate of 
the system Hamiltonian after the quench -- but rather a linear combination of its multiple 
eigenstates -- $\mathcal{L}(t)$ usually shows a complex oscillatory dependence on time $t$, 
indicating the presence of dynamical recurrences.

For the sake of comparison with the proposed simulator, once implemented on a NISQ hardware, 
the system dynamics can be computed classically. Because $[\hat{H}_{\textrm{sys}},\hat{K}_{\mathrm{tot}}]=0$, 
the system evolves within the eigensubspace of $\hat{H}_{\textrm{sys}}$ 
that corresponds to the eigenvalue $K=0$ of $\hat{K}_{\mathrm{tot}}$. Its state $|\psi(t)\rangle$ at time 
$t$ is computed here for $N=9$ qubits by combining Lanczos-type exact diagonalization~\cite{CullumWilloughbyBook}
of $\hat{H}_{\textrm{sys}}$ in a symmetry-adapted basis of the truncated Hilbert space of the system 
and an expansion of $\hat{U}_{\textrm{sys}}$ into a finite series of Chebyshev polynomials
of the first kind~\cite{TalEzer+Kosloff:84,Kosloff:94}.

\textit{Conclusions and Outlook}.-- Aiming to provide a reliable testbed for exploring 
the nonequilibrium dynamics of SP formation following an e-ph interaction quench, this paper 
presented a DQS of nonlocal e-ph interactions. Quantum circuits emulating e-ph interactions of 
Peierls- and breathing-mode types, along with a circuit generating the initial $W$-type state 
of the system, form the backbone of the envisioned simulator. To enable future comparisons 
with this simulator -- once practically realized -- both the ground-state properties of the 
considered model and its dynamics were computed classically in a numerically-exact fashion. 

The DQS proposed here will facilitate advances in the general understanding of the nonequilibrium 
properties of polaronic quasiparticles~\cite{Jager+Barnett:21}. More generally speaking, it strongly
reinvigorates the idea of using DQS to investigate the quench dynamics of quantum many-body systems~\cite{Mitra:18}, 
a research approach which is still in the stage of infancy~\cite{Smith+:19} despite the developments
of quantum hardware.

The proposed DQS complements the existing simulators of interacting fermions. Hybrid simulators, capable
of emulating both strong electron correlations and e-ph interactions, may become indispensable in  
investigating complex electronic materials. A case in point are cuprates, the parent compounds of 
high-temperature superconductors, where the two e-ph coupling mechanisms addressed here play important 
roles~\cite{Shneyder+:20}. While the currently available quantum-computing setups~\cite{Wendin:17,Bruzewicz+:19,
MorgadoWhitlock:21} (systems containing from $50$ to a few hundred qubits) belong to the realm of 
noisy intermediate-scale quantum technology~\cite{PreskillNISQ:18}, their further development (in 
terms of qubit count, quality, and connectivity), along with the optimization of DQS algorithms~\cite{Raeisi+:12,Childs+:18} 
of the kind proposed here using machine-learning-based methods~\cite{Bolens+Heyl:21,Geller+:21}, 
may allow exploration of complex many-body phenomena.
\begin{acknowledgments}
This research was supported by the Deutsche Forschungsgemeinschaft (DFG) -- SFB 1119 -- 236615297.
\end{acknowledgments}

\end{document}